\documentclass[twoside,slac_one]{revtex4}
\usepackage{graphicx}
\usepackage{fancyhdr}
\usepackage{amsmath} 
\usepackage{bm}
\usepackage{amsxtra}

\usepackage{amssymb}
\usepackage{amsthm}
\usepackage{latexsym}
\usepackage{lscape}

\begin{document}

\title{Background Rejection in the DMTPC Dark Matter Search Using Charge Signals}

%

\author{J.P.~Lopez}\email{jplopez@mit.edu}
\affiliation{Department of Physics, Massachusetts Institute of Technology; Cambridge, MA, USA}
\author{S.~Ahlen}
\affiliation{Department of Physics, Boston University; Boston, MA, USA}
\author{J.~Battat}
\affiliation{Department of Physics, Bryn Mawr College; Bryn Mawr, PA, USA}
\author{T.~Caldwell}
\affiliation{Department of Physics, Massachusetts Institute of Technology; Cambridge, MA, USA}
\author{M.~Chernicoff}
\affiliation{Department of Physics, Boston University; Boston, MA, USA}
\author{C.~Deaconu}
\affiliation{Department of Physics, Massachusetts Institute of Technology; Cambridge, MA, USA}
\author{D.~Dujmic}
\affiliation{Department of Physics, Massachusetts Institute of Technology; Cambridge, MA, USA}
\affiliation{Laboratory for Nuclear Science, Massachusetts Institute of Technology; Cambridge, MA, USA}
\author{A.~Dushkin}
\affiliation{Department of Physics, Brandeis University; Waltham, MA, USA}
\author{W.~Fedus}
\affiliation{Department of Physics, Massachusetts Institute of Technology; Cambridge, MA, USA}
\author{P.~Fisher}
\affiliation{Department of Physics, Massachusetts Institute of Technology; Cambridge, MA, USA}
\affiliation{Laboratory for Nuclear Science, Massachusetts Institute of Technology; Cambridge, MA, USA}
\affiliation{MIT Kavli Institute for Astrophysics and Space Research, Massachusetts Institute of Technology; Cambridge, MA, USA}
\author{F.~Golub}
\affiliation{Department of Physics, Brandeis University; Waltham, MA, USA}
\author{S.~Henderson}
\affiliation{Department of Physics, Massachusetts Institute of Technology; Cambridge, MA, USA}
\author{A.~Inglis}
\affiliation{Department of Physics, Boston University; Boston, MA, USA}
\author{A.~Kaboth}
\affiliation{Department of Physics, Massachusetts Institute of Technology; Cambridge, MA, USA}
\author{G.~Kohse}
\affiliation{Department of Nuclear Science and Engineering, Massachusetts Institute of Technology; Cambridge, MA, USA}
\author{L.~Kirsch}
\affiliation{Department of Physics, Brandeis University; Waltham, MA, USA}
\author{R.~Lanza}
\affiliation{Department of Nuclear Science and Engineering, Massachusetts Institute of Technology; Cambridge, MA, USA}
\author{A.~Lee}
\affiliation{Department of Physics, Massachusetts Institute of Technology; Cambridge, MA, USA}
\author{J.~Monroe}
\affiliation{Department of Physics, Royal Holloway, University of London;  Egham, Surrey, UK}
\affiliation{Department of Physics, Massachusetts Institute of Technology; Cambridge, MA, USA}
\author{H.~Ouyang}
\affiliation{Department of Physics, Brandeis University; Waltham, MA, USA}
\author{T.~Sahin}
\affiliation{Department of Physics, Massachusetts Institute of Technology; Cambridge, MA, USA}
\author{G.~Sciolla}
\affiliation{Department of Physics, Brandeis University; Waltham, MA, USA}
\author{N.~Skvorodnev}
\affiliation{Department of Physics, Brandeis University; Waltham, MA, USA}
\author{H.~Tomita}
\affiliation{Department of Physics, Boston University; Boston, MA, USA}
\author{H.~Wellenstein}
\affiliation{Department of Physics, Brandeis University; Waltham, MA, USA}
\author{I.~Wolfe}
\affiliation{Department of Physics, Massachusetts Institute of Technology; Cambridge, MA, USA}
\author{R.~Yamamoto}
\affiliation{Department of Physics, Massachusetts Institute of Technology; Cambridge, MA, USA}
\author{H.~Yegoryan}
\affiliation{Department of Physics, Massachusetts Institute of Technology; Cambridge, MA, USA}

\begin{abstract}
The Dark Matter Time Projection Chamber (DMTPC) collaboration is developing low-pressure gas TPC detectors for measuring WIMP-nucleon interactions.  Optical readout with CCD cameras allows for the detection for the daily modulation in the direction of the dark matter wind, while several charge readout channels allow for the measurement of additional recoil properties.  In this article, we show that the addition of the charge readout analysis to the CCD allows us too obtain a statistics-limited 90\%~C.L. upper limit on the $e^-$ rejection factor of $5.6\times10^{-6}$ for recoils with energies between 40 and 200~keV$_{\mathrm{ee}}$.  In addition, requiring coincidence between charge signals and light in the CCD reduces CCD-specific backgrounds by more than two orders of magnitude.
\end{abstract}

\maketitle

\thispagestyle{fancy}


\section{Introduction\label{sec:intro}}

The DMTPC collaboration is developing low-pressure gas time projection chambers (TPCs) with sensitivity to recoil directions to search for weakly interacting massive particles (WIMPs) in the galactic dark matter halo.  The interaction signature is a low-momentum nuclear recoil that leaves a trail of ionization inside the target gas volume of the detector.  DMTPC detectors use 50 to 100 Torr CF$_4$ gas to measure spin-dependent interactions between WIMPs and $^{19}$F nuclei. Fluorine recoils with energies between 50 and 200~keV, the target energy range for DMTPC dark matter analyses, have typical ranges of 1 to 2~mm, long enough to determine the direction of the recoil. The large daily modulation of the direction of the dark matter wind is expected to be distinct from all known backgrounds and will allow such a detector to distinguish between a dark matter signal and background events with a small number recoils~\cite{Ahlen:2009ev,Spergel}. 

Electronic recoils are an important background in most dark matter direct detection experiments.  DMTPC detectors use charge-coupled device (CCD) cameras to obtain fine pixelation of the TPC readout plane and as a result are unable to detect most electronic recoils.  Several charge readout channels detect such recoils. Pulse shapes are use to discriminate between nuclear recoils and electrons.  The charge readout channels provide independent measurements of recoils in the detector and are also used to improve our discrimination of CCD-related backgrounds. This article describes the preliminary results of a short run taken to evaluate the capabilities of the charge readout channels to reject both electronic recoils and CCD backgrounds.
\section{DMTPC Detector Design}

\begin{figure}
  \setlength{\unitlength}{0.85mm}
\begin{picture}(140,120)
\put(84,113){\framebox(14,6) {CCD}} 
\put(91,113){\line(-1,-2){3}}
\put(91,113){\line( 1,-2){3}}
\put(95,109){\small lens}
\multiput(50,27)(0,7){10}{\framebox(10,2)}
\multiput(122,27)(0,7){10}{\framebox(10,2)}
\put(25,92.1){\line(1,0){35}}
\multiput(60,92.1)(1,0){68}{\line(1,0){0.5}}
\put(15,88){\framebox(10,8){-HV}}
\put(136,93){\small cathode}
\put(136,90){\small mesh}
\put(25,17){\line(0,1){10}} \put(25,27){\line(-1,-1){5}} \put(25,17){\line(-1,1){5}} \put(20,22){\line(-1,0){5}}
\put(35,16){\framebox(100,90)}
\put(50,107){\footnotesize Vacuum Vessel}
\put(25,22){\line(1,0){25}} \put(50,22){\circle*{1}} \multiput(50,22)(1,0){82}{\line(1,0){0.5}} 
\put(4,14){ \footnotesize Route2Electronics}
\put(4,10){ \footnotesize HS-AMP-CF-2nF}
\put(4,6){\footnotesize mesh}
  \multiput(40,30)(0,7){9}{\framebox(2,4)}
  \multiput(41,30)(0,7){9}{\line(0,-1){3}}
  \multiput(41,28.5)(0,7){9}{\line(1,0){9}}
  \put(41,92.1){\line(0,-1){2}}
  \put(41,30){\line(0,-1){7.5}}
  \put(41.15,22){\oval(1,1)[r]}
  \put(38,12){\line(1,0){6}}\put(39,11){\line(1,0){4}} \put(40,10){\line(1,0){2}}
  \put(40,16.5){\framebox(2,4)} 
  \put(41,16.5){\line(0,-1){4.5}} 
  \put(42,16){ \footnotesize R}
  \multiput(36,31)(0,7){9}{\footnotesize R}
\put(50,18){\framebox(20,2)}  
\put(60,18){\line(0,-1){3}} 
\put(58,15){\line(1,0){4}} \put(58,14){\line(1,0){4}}
\put(60,14){\line(0,-1){3}} 
\put(55,11){\line(1,0){10}} \put(55,11){\line(1,-1){5}}  \put(65,11){\line(-1,-1){5}}  \put(60,6){\line(0,-1){6}}
\put(60,12.5){\line(1,0){7}} \put(60,4.5){\line(1,0){7}} 
\put(67,12.5){\line(0,-1){4}} \put(67,4.5){\line(0,1){3}}
\put(65,8.5){\line(1,0){4}} \put(65,7.5){\line(1,0){4}}
\put(70,8.5){\footnotesize Cremat}
\put(70,5.5){\footnotesize CR-112}
\put(70,2.5){\footnotesize (veto)}
\put(136,23){\small grounded}
\put(136,20){\small mesh}
\put(105,13.5){\footnotesize anode plane, +HV}
\put(112,18){\framebox(20,2)}  
\put(71,18){\framebox(40,2)}
\put(90,18){\line(0,-1){3}} 
\put(88,15){\line(1,0){4}} \put(88,14){\line(1,0){4}}
\put(90,14){\line(0,-1){3}} 
\put(85,11){\line(1,0){10}} \put(85,11){\line(1,-1){5}}  \put(95,11){\line(-1,-1){5}}  \put(90,6){\line(0,-1){6}}
\put(60,14){\line(0,-1){3}} \put(90,12.5){\line(1,0){7}} \put(90,4.5){\line(1,0){7}} 
\put(97,12.5){\line(0,-1){4}} \put(97,4.5){\line(0,1){3}}
\put(95,8.5){\line(1,0){4}} \put(95,7.5){\line(1,0){4}}
\put(100,8.5){\footnotesize Cremat}
\put(100,5.5){\footnotesize CR-113}
\put(100,2.5){\footnotesize (anode)}
\put(130,100){\vector(-3,-1){45}}
\put(120, 102){$^{241}$Am}
\put(60,97){75 Torr CF$_4$}
\put(115,97){$\alpha$}
\put(133, 102){\line(-3,-1){5}}
\put(133.5, 100.5){\line(-3,-1){5}}
\put(97,87){\circle*{1}}  \put(100,91){\circle*{1}} 
\put(98,88){\circle*{1}}  \put(100,88){\circle*{1}}
\put(98,91){\circle*{1}}  \put(90,87){\circle*{1}}
\put(95,89){\circle*{1}}  \put(93,86){\circle*{1}}
\put(88,85){\circle*{1}}  \put(99,89){\circle*{1}}
\put(90,30){\small ionization electrons}
\put(97,24){\circle*{1}}  \put(100,28){\circle*{1}} 
\put(98,25){\circle*{1}}  \put(100,25){\circle*{1}}
\put(98,28){\circle*{1}}  \put(90,24){\circle*{1}}
\put(95,26){\circle*{1}}  \put(93,23){\circle*{1}}
\put(99,26){\circle*{1}}
%
\put(87,20){\large $\star$}
\end{picture}
\caption{A schematic of the detector: the drift field is created by a cathode mesh, field-shaping rings attached to a resistor chain and a ground mesh. 
Electrons from ionization left by a recoiling nucleus drift down to the grounded mesh. The high-field amplification region is formed by the grounded
mesh and the anode plane.  The grounded mesh is read out with a fast amplifier and the veto and anode are readout with charge-sensitive preamplifiers. 
Scintillation light from the amplification region is recorded with the CCD camera, which is located outside the vacuum vessel. The CCD is attached to a lens focused on the anode plane and looks through a viewport on the top of the vessel.
\label{schematic}}
\end{figure}
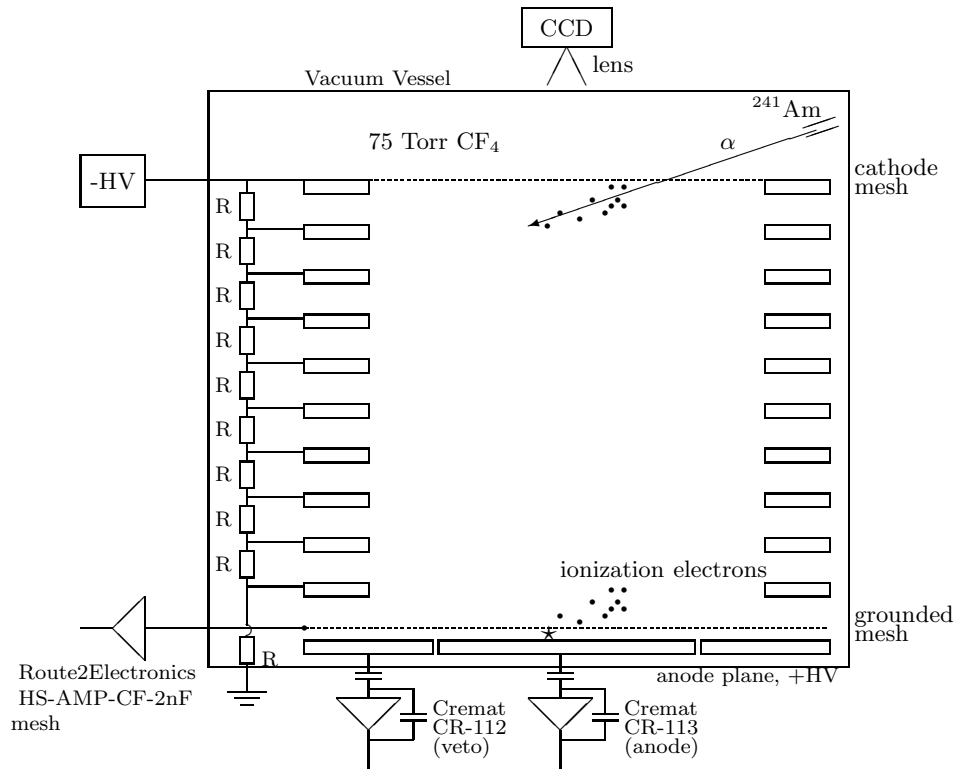

The data in this article was taken in a laboratory in Cambridge, Massachusetts with a small prototype DMTPC detector.  The basic design of a DMTPC detector is described in ~\cite{Dujmic:2008ut}.  The detector, shown in more detail in Fig. \ref{schematic}, has a field cage with a drift length of 10~cm.  The two ends of the field cage are bounded by wire meshes, with the cathode mesh held at -1.2~kV and the other held at ground.  The drift cage is created by a series of copper rings with an inner diameter of 26.7~cm held 1~cm apart to create a nearly uniform electric field.  An amplification gap is created by separating the grounded mesh from a plate of copper-clad G10 using several 440~$\mu$m diameter dielectric wires.  Electron avalanches with gains of $10^4$ to $10^5$ are achieved with this design.  Scintillation light from the avalanches is read out by an Apogee Alta U6 CCD camera, while several charge readout channels read the current and integrated charge signals from the grounded mesh and the anode.  The high gain is necessary to counteract the large reduction in scintillation light incident on the CCD due to the lens acceptance.

The CCD camera uses a Kodak KAF-1001 1024 by 1024 pixel CCD. Signals are digitized and read out in 4 by 4 pixel bins.  A lens focuses light from the anode plate onto the CCD so that each of these bins views a 650~$\mu$m by 650~$\mu$m region of the anode.  The total area imaged by the CCD is 16.7~cm by 16.7~cm, roughly half of the anode.  The central part of the anode is viewed, so the field cage rings are not visible.  A $4.44\pm0.04$ MeV $^{241}$Am source, calibrated in \cite{HaykThesis}, was placed inside the field cage to calibrate CCD energy measurements.  The mean value of the source distribution yields a CCD energy calibration of $13.1\pm0.1$ analog-to-digital units (ADU) per keV of energy deposited.  This source was then removed for the remaining data.  Nearly all of the energy loss of $\alpha$ particles at this energy is due to ionization \cite{SRIM}, so this calibration is very similar to the electron-equivalent ionization energy. Measured energies used in this article are in terms of $\alpha$-equivalent ionization energy loss, called keV$_{\mathrm{ee}}$ here.

The anode plane is split into an outer ring 1~cm wide (veto channel) to identify tracks passing near the edge of the field cage and an inner region (anode channel). The anode channel is fed through a Cremat CR-113 charge-sensitive preamplifier (CSP), while the veto channel is fed through a Cremat CR-112 CSP. The CR-113 and CR-112 CSPs have gains of 1.5~mV/pC and 15~mV/pC, respectively, and both have a decay constant of 50~$\mu$s. The typical rise time of a signal pulse is roughly 1~$\mu$s, so these accurately amplify the interated charge signal of a nuclear recoil. 

The grounded mesh signal is amplified by a Route2Electronics HS-AMP-CF current-sensitive preamplifier, with a gain of 80 and a rise time of approximately 1~ns. The pulse shape of the current signal provides information about recoil geometry and is used to reject the signals of most electronic recoils and minimum ionizing particles. In the avalanche of a single electron drifting into the amplification region, the ionization occurs near the anode plane. Electrons drift toward the anode plane, creating a large current peak that lasts for less than 1~ns. The ions left over from ionization in the amplification gap drift toward the grounded mesh at a much slower velocity than the electrons, so the current signal is smaller but the integrated charge is larger from the higher drift time of the ions.  

\begin{figure}
\includegraphics[width=80mm]{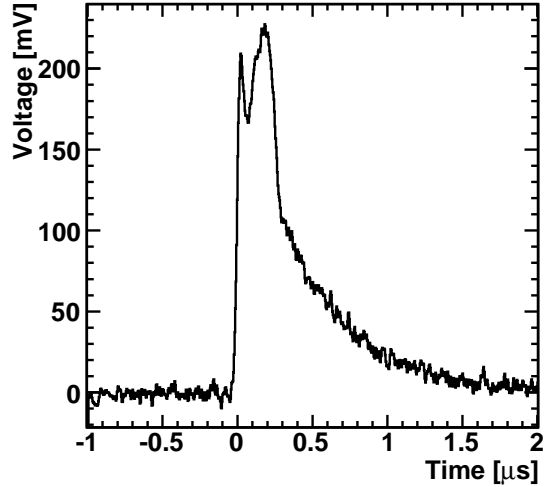}
\caption{The mesh (current) signal of a low energy $\alpha$ track. This pulse shape is very similar to those obtained for fluorine and carbon nuclear recoils.  An initial electron peak is seen shortly after the sharp rising edge. A second distinct peak from the drifting ions is present following that peak.}
\label{ExAl}
\end{figure}

\begin{figure}
\includegraphics[width=80mm]{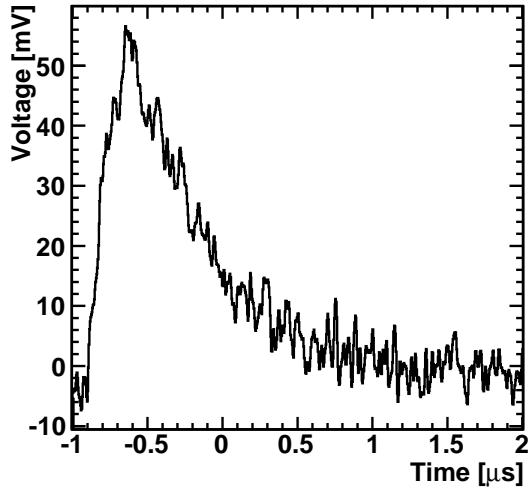}
\caption{The mesh (current) signal of an electronic recoil. Unlike the $\alpha$ signal in Fig. \ref{ExAl}, there is only one clear peak in this signal.  This is because the higher $\Delta z$ of electronic recoils causes the drifting electrons to reach the amplification region too far apart to add into a single peak.}
\label{ExEl}
\end{figure}

In the drift region, nuclear recoils are compact. The small range along the drift direction ($\Delta z$) corresponds to a time spread between the initial and final avalanches from primary ionization electrons is no more than a few tens of nanoseconds. The resulting current signal has clearly identifiable peaks from the electrons and ions left by ionization from the avalanches, seen in an $\alpha$ signal in Fig. \ref{ExAl}. Electrons and minimum ionizing particles, such as that seen in Fig. \ref{ExEl}, leave small amounts of ionization throughout a large region of the detector. These have a much longer typical $\Delta z$, and the electron and ion signals merge into a single peak with a much longer rise time than a nuclear recoil of the same energy. This difference in pulse shape is used to differentiate electrons from nuclear recoils and $\alpha$ particles. 

The signals from the charge channels are digitized using AlazarTech ATS860 8-bit digitizers. The digitizer is triggered on either the anode or the mesh channel, and the trigger level is set to achieve high detection efficiency for energies above 30 keV$_{\mathrm{ee}}$.  The waveforms are 49.152~$\mu$s in length, with the pre-trigger samples comprising one third of the waveform.   A single event consists of the image from a 1~s exposure of the CCD and the waveforms from any triggers of the charge channels during the CCD exposure. A camera shutter is not used for the CCD, so the true CCD exposure includes an additional 0.2-0.3~s due to readout time.  

\section{Event Reconstruction and Analysis}

The reconstruction of recoils in CCD images is described in detail in \cite{10LPaper}.  This article uses a simpler set of selection cuts than were included in the analysis in \cite{10LPaper} in order to maximize the efficiency for nuclear recoils.  CCD related backgrounds are identified by searching for potential recoils in which the peak apparent light signal in a single pixel is either very high or contains too much of the total measured light.  Events reaching the edge of the image are likely to be from edge effects on the image noise during reconstruction or are partially contained tracks. The peak pixel values of the measured recoil candidates must also be high enough to obtain a reasonable range estimate.

The anode and veto signal reconstructions use the same algorithms.  The baseline is characterized by its mean and standard deviation using a subset of pre-trigger samples.  The peak value and time are recorded for each waveform. The positions of the baseline crossings nearest to the peak are calculated and the times where the waveform reaches 10\%, 25\%, 50\%, 75\%, and 90\% of the peak value on the rising and falling edges of the pulse are recorded.

For the mesh channel, the baseline, peak time and peak height are calculated in the same way as for the anode and veto channels.  The times and heights of the electron and ion peak and the minimum point between the two peaks are also calculated.  The same set of times along the rising and falling edges as for the other channels are calculated as well.  For mesh pulses, the rising edge is defined as the region between the initial baseline crossing and the electron peak, while the falling edge is the region between the ion peak and the final baseline crossing. The pulse integral is calculated to give an additional estimate of the energy.

Noise removal cuts identify events where the baseline characteristics and measured pulse rise and fall times deviate far from the expected range for nuclear recoils. Pulses not fully contained within the waveform and pulses exceeding the maximum input for the digitizer settings are identified and removed as well. Ionization caused by recoils near the veto region is identified by a large peak pulse height in the veto channel or by a long rise time (greater than 800~ns).  Events where the mesh and veto signals are not well correlated in time occur rarely but are rejected as this is indicative of signal pileup from several recoils occurring in a single waveform. Nuclear recoils are then identified by a short rise time and large electron and ion peak heights when compared to the total recoil energy. 

These cuts were determined by placing an $^{241}$Am $\alpha$ source above the cathode mesh of the field cage, seen in Fig. \ref{schematic}, so that we only measure the last few tens to hundreds of keV of the $\alpha$ track.  Alpha particles are slightly longer than nuclear recoils and their position in the drift region maximizes the diffusion. As a result, the mesh signals in $\alpha$ data have longer rise times and less prominent peaks than would be seen for fluorine and carbon recoils and are useful for setting conservative cuts to remove background events.

Finally, the recoils measured in the CCD are matched to the corresponding charge signal by comparing the energy measured by the CCD to the energy measured by the anode channel.  The $\alpha$ data was used to determine the best fit line between the two channels,  $E_{\mathrm{anode}}[\mathrm{mV}] = (3.07 \pm 0.04) + (0.01916 \pm 0.00002)E_{\mathrm{CCD}}[\mathrm{ADU}]$. Charge and CCD signals are accepted if the measured anode energy differs from the estimated value from the CCD energy by less than 8.5~mV.

\section{Electronic Recoils\label{sec:elrej}}

Because electronic recoils have a very small stopping power in low-pressure gases, the CCD analysis is typically blind to these events.  Not enough light is detected by any individual CCD pixel for the recoil to be detectable above the background noise level.  The charge channels, however, measure the full recoil signal and are able to measure electronic recoils.  Because these are much more common than nuclear recoils, it is important to be able to eliminate as many of these as possible so that it is possible to discover a small number of signal events.  By requiring coincidence between charge and light signals, we are able to remove electronic recoils with much better efficiency than with just the CCD or just the charge readout.  The ability of the charge analysis to reject electronic recoils is measured by placing a $^{137}$Cs $\gamma$ source inside the detector, similar to the placement of the $\alpha$ source in Fig. \ref{schematic}.  The 660 keV gammas from the source generate an average of 27 electronic recoils with $40$~keV$_{\mathrm{ee}}<E<200$~keV$_{\mathrm{ee}}$ per one second exposure.  After removing events where a spark occurred on the anode, 14.18 hours of data are used in this setup.  A separate run with no sources inside the chamber was taken to measure the background event rates.  In this setup, 9.05 hours of data are used.

After removing the estimated number of background events, the source provided $602600\pm350$ recoils with $40$~keV$_{\mathrm{ee}}<E<200$~keV$_{\mathrm{ee}}$, but only $200\pm9$ events above the background pass the charge cuts, giving a rejection factor of  $3.32\pm0.15\times10^{-4}$ for electronic recoils using just the charge analysis.  Adding the CCD analysis and matching charge signals to CCD signals, only five events pass all cuts in the $^{137}$Cs data set with an expected background of $4.7\pm2.7$. These events have ranges and peak CCD pixel values consistent with nuclear recoils rather than what is expected of an electronic recoil event.   The five passing events can then be rejected as electronic recoil candidates with very high confidence, leaving no signal events from the $^{137}$Cs data.  The distribution of light from recoils left by the source is measured from the mean of many images and used to determine that 68\% of all charge events occurred in the region imaged by the CCD.  This is used to obtain a statistics-limited 90\% confidence level upper limit on the electronic recoil rejection factor of the combined CCD and charge analysis of $5.6\times10^{-6}$.  This was taken with a much higher event rate than is seen in WIMP searches and therefore provides a very conservative estimate of ability of a DMTPC detector to reject electron events.

\section{CCD-Related Backgrounds}

\begin{figure}[ht]
\includegraphics[width=80mm]{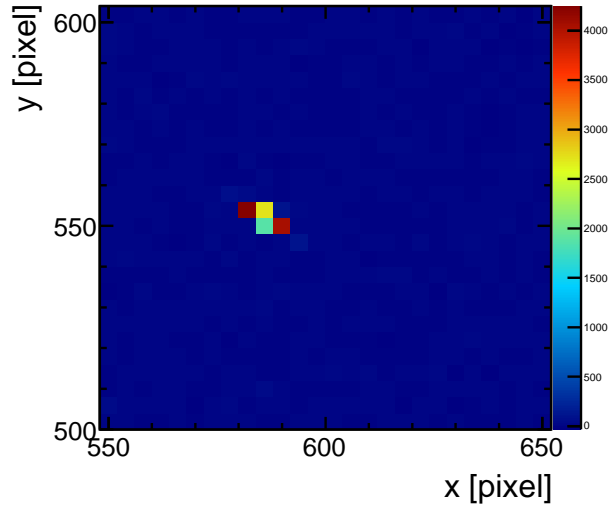}
\caption{An example of an interaction inside the CCD silicon during a run using a $^{252}$Cf source. }
\label{CCDtrack}
\end{figure}

\begin{figure}[ht]
\includegraphics[width=80mm]{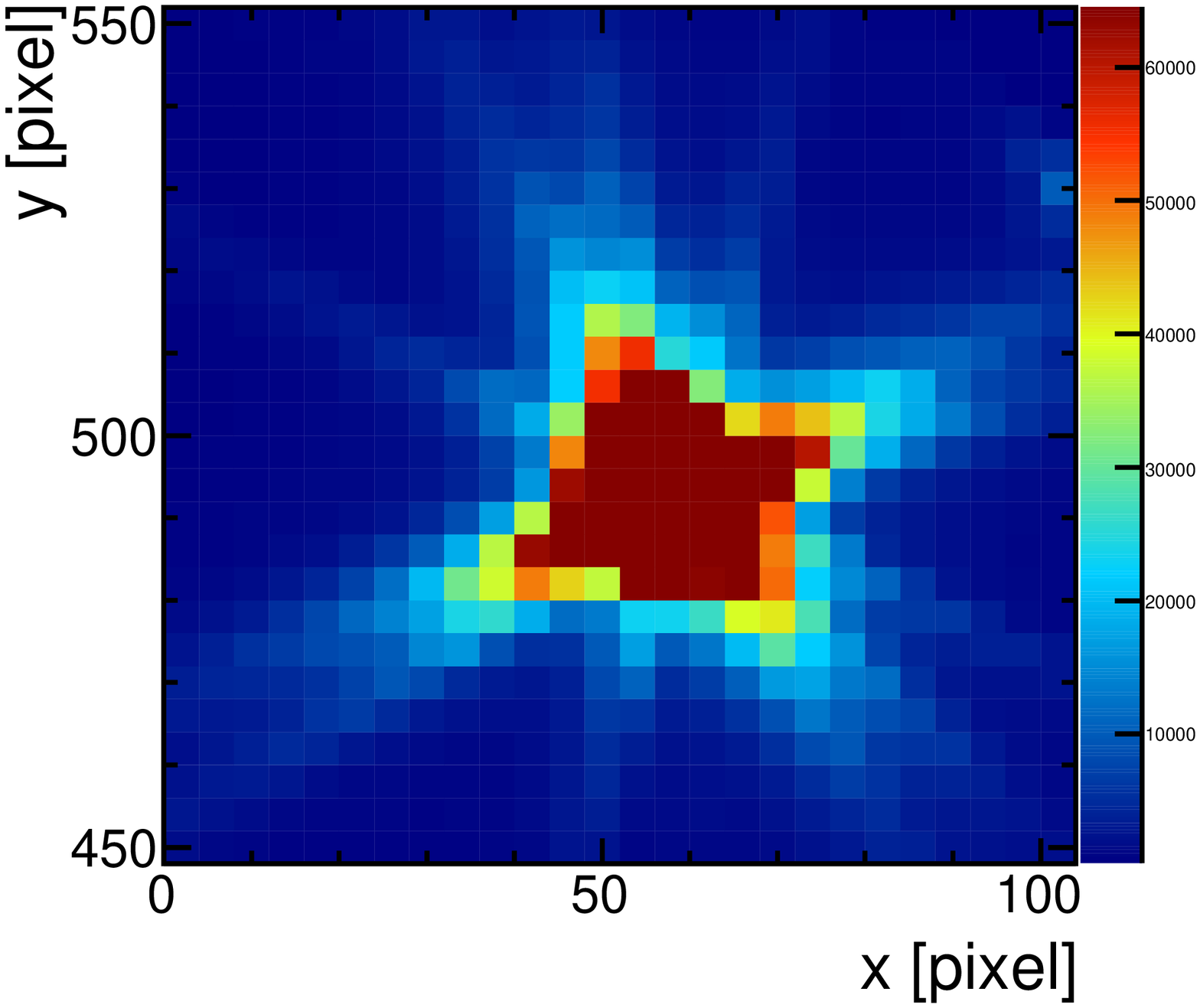}

\includegraphics[width=80mm]{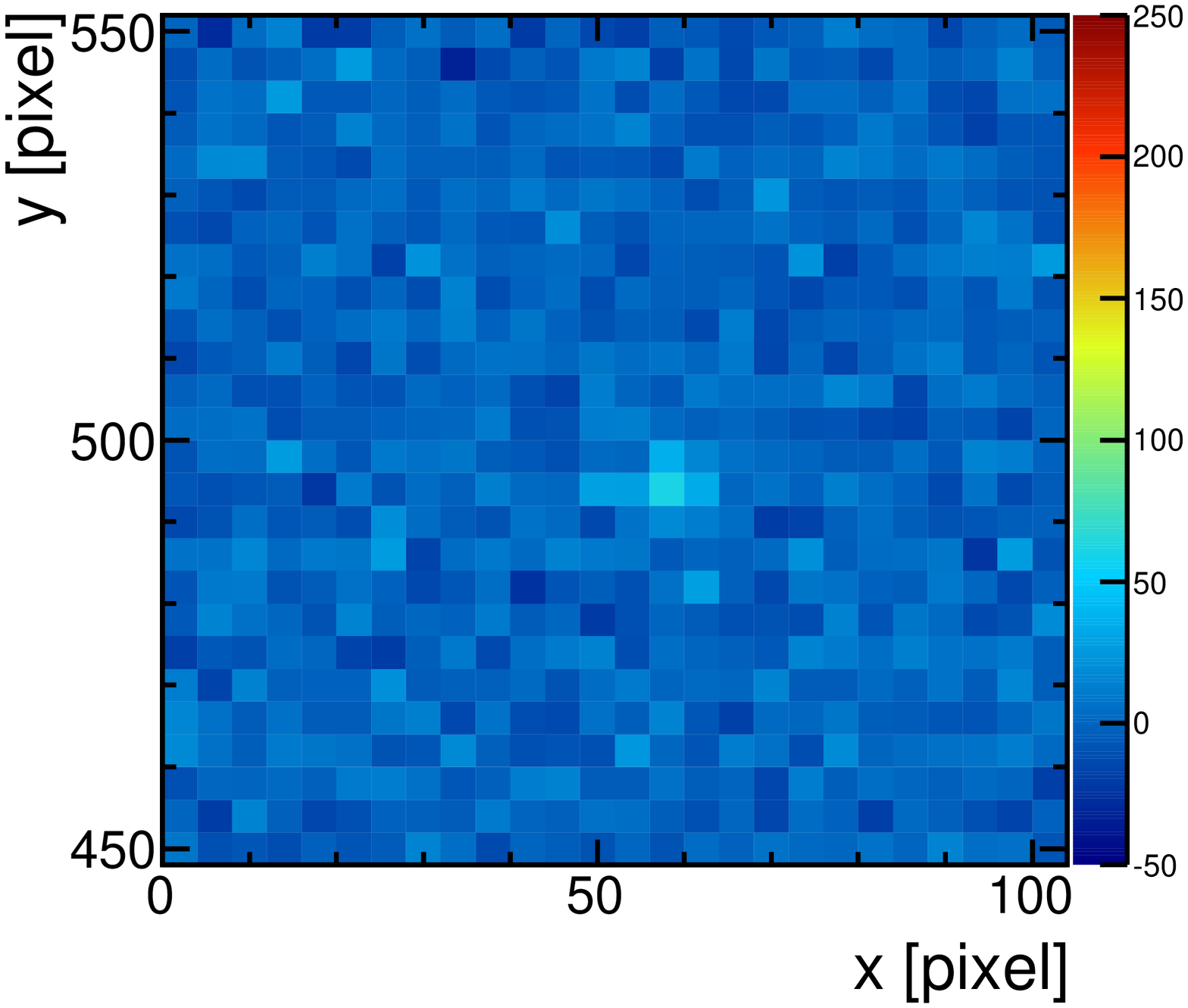}
\caption{Top: A spark occurring in a position visible to the CCD. Bottom: A faint RBI occurring in the same spot 1~s after the spark.  Some RBIs can last for several minutes.}
\label{RBIfig}
\end{figure}

A number of backgrounds due to CCD effects are also seen.  Hot pixels and interactions in the CCD chip typically look very different from nuclear recoils but are very common and must be removed from analysis.  A CCD interaction event is shown in Fig. \ref{CCDtrack}.  Residual bulk images (RBIs) are a background that occurs when charge is trapped at an interface between different layers of a front-illuminated CCD.  The charge must dissipate by thermal diffusion over several minutes~\cite{CCDbook}.  This often happens following a spark inside the detector, when a great deal of light is concentrated on a small part of the CCD.  An example of an RBI event is shown in Fig. \ref{RBIfig}.  Residual bulk images can often mimic the signal of a low energy recoil but are typically identifiable because they persist in the same location for an extended period of time.

Background $\alpha$ decay events are another background seen in DMTPC detectors.  These typically occur near the field cage rings so only a small fraction of the track is visible to the CCD.  Because the CCD shutter is not used, the readout geometry of the CCD shifts the light of an event occurring during readout from its true position.  The apparent position of an $\alpha$ decay occurring during readout is sometimes shifted so that the part of the track visible to the CCD appears near the center of the image, where it might be identified as a nuclear recoil. These out-of-time events are very difficult to identify with a CCD-only analysis.  The acquisition of charge data is suspended during CCD readout, so no charge signal is recorded .

The CCD-related backgrounds do not have matching charge signals. By requiring coincident charge and light signals, we can further suppress the types of events that are identifiable with CCD analysis and we can remove events that can only be identified with the charge analysis. Using the source free data set described in Section \ref{sec:elrej}, the charge readout analysis and charge-light energy matching is used to determine the rejection power of the charge readout for the different classes of CCD backgrounds Table \ref{ccdtable} summarizes the results.  The charge readout reduces the RBI, hot pixel and CCD interaction background by more than two orders of magnitude.  All background $\alpha$ events crossing the edge of the image are rejected as well.  Of the 14 events with $25$~keV$_{\mathrm{ee}}<E_{\mathrm{CCD}}<500$~keV$_{\mathrm{ee}}$ passing all CCD cuts, only 5 have a corresponding charge signal. These are candidate nuclear recoil events. The remaining 9 candidate recoils are likely out-of-time $\alpha$ decay events. 

\begin{table}[ht]
\caption{Preliminary results for the number of CCD events of different types found, before and after applying charge cuts, 25 keV$_{\mathrm{ee}}$ $< E_{CCD} < 500$ keV$_{\mathrm{ee}}$.  The CCD events are classified using the CCD analysis.  Reduction is the percentage of events of each type that fail the charge analysis.  The RBI and hot pixel/CCD Si recoil events are false coincidences between low energy charge signals and CCD artifacts with low apparent light signals. }
\begin{tabular}{|l|l|l|l|}
\hline
\textbf{Event type}        &     \textbf{Before} &  \textbf{After}      & \textbf{Reduction [\%]}\\
\textbf{from CCD analysis} & \textbf{charge cuts}&\textbf{charge cuts}& \\
\hline
RBI &1320&9 & $99.3\pm0.2$\\
\hline
Hot pixel/CCD Si Recoil&1287 &9 &$99.3\pm0.2$ \\
\hline
Edge Crossing ($\alpha$) &15 &0 & $100_{-7}^{+0}$\\
\hline
Nuclear recoil/Out of time $\alpha$ &14 &5 &$60\pm10$\\
\hline
All Tracks &2636 & 23 &$99.1\pm0.2$\\
\hline
\end{tabular}

\label{ccdtable}
\end{table}

\section{Conclusions}

The charge readout study presented here shows that even a simple analysis can significantly reduce both electronic recoil backgrounds and CCD-related backgrounds. We have demonstrated the ability of the charge readout analysis to reject electronic recoils with a rejection factor of $3.32\times10^{-4}$ for energies between 40 and 200 kev$_{\mathrm{ee}}$.  This reduces the rate of charge events enough to accurately match them to their corresponding CCD signal.  For the combined CCD and charge analysis, a 90\% C.L. upper limit on the electronic recoil rejection factor of $5.6\times10^{-6}$ is obtained for the same energy range.  In WIMP searches, the rejection power for electronic recoils is expected to be much better than this limit due to having far fewer electronic recoils accumulated in each CCD exposure.

We have also demonstrated that the addition of charge readout analysis can reduce CCD related backgrounds by more than two orders of magnitude.  This study also shows that with the addition of the charge readout analysis, DMTPC detectors are now able to identify out-of-time events from $\alpha$ decays occurring during CCD readout.  This category of event had previously been unidentifiable with CCD-only analyses. The charge readout systems have been added to other detectors used by the DMTPC collaboration, and the analysis will be incorporated into future WIMP searches to improve background rejection and to gain more information about nuclear recoil candidates.


\begin{acknowledgments}
The DMTPC collaboration would like to acknowledge support by the U.S. Department of Energy (grant number
DE-FG02-05ER41360), the Advanced Detector Research Program of
the U.S. Department of Energy (contract number 6916448), as well as the
Reed Award Program, the Ferry Fund, the Pappalardo Fellowship program,
the MIT Kavli Institute for Astrophysics and Space Research, the MIT
Bates Research and Engineering Center, and the Physics Department at the
Massachusetts Institute of Technology. We would like to thank Mike
Grossman for valuable technical assistance.

\end{acknowledgments}

\bigskip 


\end{document}